\documentclass[conference,a4paper]{IEEEtran}

\IEEEoverridecommandlockouts

\usepackage{graphicx}
\usepackage{amsmath}
\usepackage{amssymb}
\usepackage{color}
\usepackage{multirow}
\usepackage{tabularx}
\usepackage{balance}
\usepackage[numbers]{natbib}
\usepackage{url}

\usepackage{algorithm}
\usepackage{algpseudocode}

\DeclareMathOperator*{\argmin}{arg\,min}
\algnewcommand\algorithmicinput{\textbf{Input:}}
\algnewcommand\INPUT{\item[\algorithmicinput]}
\algnewcommand\algorithmicoutput{\textbf{Output:}}
\algnewcommand\OUTPUT{\item[\algorithmicoutput]}

\usepackage{mathtools}

\newcommand{\Fig}[1]{Fig.~\textup{\ref{#1}}}
\graphicspath{{figures/}}
\usepackage{subcaption}

\newtheorem{theorem}{Theorem}

\newtheorem{remark}{Remark}

\newtheorem{statement}{Statement}

\begin{document}

\title{Achievability Bounds for T-Fold Irregular Repetition Slotted ALOHA Scheme in the Gaussian MAC\thanks{The research was carried out at Skoltech and supported by the Russian Science Foundation (project no. 18-19-00673).}}

\author{
  \IEEEauthorblockN{Anton Glebov\IEEEauthorrefmark{1}, Nikolay Matveev\IEEEauthorrefmark{3}, Kirill Andreev\IEEEauthorrefmark{1}, Alexey Frolov\IEEEauthorrefmark{1} and Andrey Turlikov \IEEEauthorrefmark{3}}
	
 \IEEEauthorblockA{\small \IEEEauthorrefmark{1} Skolkovo Institute of Science and Technology, Moscow, Russia
    }
 \IEEEauthorblockA{\small \IEEEauthorrefmark{3} State University of Aerospace Instrumentation, St. Petersburg, Russia
    }

  {anton.glebov@skolkovotech.ru, n.matveev@vu.spb.ru, k.andreev@skoltech.ru, al.frolov@skoltech.ru, turlikov@vu.spb.ru}
      
}


%


\maketitle
\begin{abstract}
We address the problem of uncoordinated massive random-access in the Gaussian multiple access channel (MAC). The performance of low-complexity $T$-fold irregular repetition slotted ALOHA (IRSA) scheme is investigated and achievability bounds are derived. The main difference of this scheme in comparison to IRSA is as follows: any collisions of order up to $T$ can be resolved with some probability of error introduced by noise. In order to optimize the parameters of the scheme we combine the density evolution method (DE) proposed by G.~Liva and a finite length random coding bound for the Gaussian MAC proposed by Y.~Polyanskiy. As energy efficiency is of critical importance for massive machine-type communication (mMTC), then our main goal is to minimize the energy-per-bit required to achieve the target packet loss ratio (PLR). We consider two scenarios: (a) the number of active users is fixed; (b) the number of active users is a Poisson random variable.

\end{abstract}
\section{Introduction}
Existing wireless networks are designed with the goal of increasing a spectral efficiency in order to serve human users. Next generation of wireless networks will face a new challenge in the form of machine-type communication. The main challenges are as follows: (a) huge number of autonomous devices connected to one access point, (b) low energy consumption, (c) short data packets. This problem has attracted attention of 3GPP standardization committee under the name of mMTC (massive machine-type communication). Numerous solutions can be found in 3GPP proposals, we mention here only three main candidates: multi-user shared access (MUSA, \cite{yuan2016non}), sparse coded multiple access (SCMA, \cite{nikopour2013sparse}) and resource shared multiple access (RSMA, \cite{3gpp.R1-164688, 3gpp.R1-164689}). Unfortunately, due to the lack of implementation details it is difficult to answer, how good these candidates are.

This paper deals with construction of low-complexity random coding schemes for the Gaussian MAC with equal-power users. In interest of reducing hardware complexity and improving energy efficiency we focus on “grant-free” transmission (3GPP terminology), which means that active users transmit their data without any prior communication with the base station. The main goal is to minimize the energy-per-bit spent by each of the users. 

We continue the line of work started in \cite{polyanskiy2017perspective, OrdentlichPolyanskiy2017}. In \cite{polyanskiy2017perspective} the bounds on the performance of finite-length codes for Gaussian MAC are presented. In \cite{OrdentlichPolyanskiy2017} Ordentlich and Polyanskiy describe the first low-complexity coding paradigm for Gaussian MAC. The solution is based on $T$-fold slotted ALOHA (SA) and carefully constructed concatenated codes, which are required in order to resolve the collisions. Unfortunately, the required energy-per-bit for this scheme is too far from the bound \cite{polyanskiy2017perspective}. We note, that $T$-fold SA is a compromise in between conventional ($1$-fold) SA, in which only single-user decoding is possible (and thus this scheme has the lowest complexity possible), and joint decoding of all active users, which transmit simultaneously with use of the same (randomly generated) codebook (exponential in the codelength complexity). In $T$-fold SA scheme the code is only required to resolve the collisions of order up to $T$, where $T$ is a relatively small value.

In this paper we investigate the potential capabilities of $T$-fold IRSA. $1$-fold IRSA (\cite{Liva, Pfister}) is known to significantly outperform $1$-fold SA for noiseless collision channel. We note, that $T$-fold IRSA is also known in the literature under the name IRSA with multipacket reception (IRSA-MPR). An analysis of IRSA-MPR over the noiseless collision channel was conducted in \cite{MPR_schlegel}. In \cite{MPR_stefanovich} the converse bound was given. We also note, that the problem considered here is close to so-called IRSA with capture effect \cite{Capture}, when sufficiently strong signals may be decoded in a slot (the authors assume, that users have different path loss coefficients). In these papers the authors aim to maximize the throughput of the resulting scheme. Our main goal is to measure the energy efficiency of $T$-fold IRSA in Gaussian MAC, as this parameter is of critical importance for mMTC scenario. We were able to find the only paper \cite{Vem2017}, in which this question was addresses. Here we present the improvement of the bounds from \cite{Vem2017}. In order to optimize the parameters of the scheme we combine the density evolution method (DE) proposed in \cite{Liva} and a finite length random coding bound for the Gaussian MAC proposed in \cite{polyanskiy2017perspective}.


Our contribution is as follows. We derive achievability bounds for $T$-Fold IRSA in the Gaussian MAC. We consider two scenarios: (a) the number of active users is fixed; (b) the number of active users is a Poisson random variable. To the best of the authors' knowledge, the bounds presented in this paper are the best achievability bounds for low-complexity random coding schemes for the Gaussian MAC.

\section{System model}

Let us describe the system model. There are $K_\text{tot} \gg 1$ users, of which only $K$ are active in each time instant. Communication proceeds in a frame-synchronized fashion (this can be implemented with use of beacons). The length of each frame is $N$. Each active user has $k$ bits to transmit during a frame.

Let us describe the channel model
\begin{equation*}
\mathbf{y} = \sum_{i=1}^{K_\text{tot}} s_i \mathbf{x}_i + \mathbf{z},   
\end{equation*}
where $\mathbf{x}_i \in \mathbb{R}^n$ is a codeword\footnote{We will make no difference between terms ``codeword'' and ``packet''} transmitted by the $i$-th user, $s_i$ is an activity indicator for the $i$-th user, i.e. $s_i = 1$ if the $i$-th user is active and $s_i = 0$ otherwise. $\mathbf{z} \sim \mathcal{N}(\mathbf{0}, \mathbf{I})$ is an additive white Gaussian noise (AWGN). Following \cite{polyanskiy2017perspective} we assume all the users to use the same message set $[M] \triangleq \{1, \ldots, M\}$ and the same codebook $\mathcal{C} = \{\mathbf{x}(\omega)\}_{\omega=1}^M$ of size $M$. Let $\omega_i$ denote the message of the $i$-th user. To transmit the message $\omega_i$ the user will use a codeword $\mathbf{x}_i = \mathbf{x}(\omega_i)$. We require in addition that $||\mathbf{x}(\omega)||^2_2 \leq NP$, which means a natural power constraint. 

Decoding is done up to permutation of messages. We only require the decoder to output a set $\mathcal{L}(\mathbf{y}) = (\omega_1, \omega_2, \ldots, \omega_K) \in [M]^K$. Thus in accordance to \cite{polyanskiy2017perspective} we decouple the user identification problem and the data transmission problem. The probability of error (per user) is defined as follows (see \cite{OrdentlichPolyanskiy2017})
\[
p_e = \max\limits_{|(s_1, s_2, \ldots, s_{K_\text{tot}})| = K} \frac{1}{K} \sum\limits_{i=1}^{K_\text{tot}} s_i \Pr(W_i \not\in \mathcal{L}(\mathbf{y})).
\]

It is clear, that the probability depends only on the messages, that were sent to the channel. Thus we can calculate it as follows
\[
p_e = \frac{1}{K} \sum\limits_{i=1}^{K} \Pr(W_i \not\in \mathcal{L}(\mathbf{y})),
\]
where $W_i$ is the $i$-th message.

Let us emphasize the main differences from the classical setting. Almost all well-known low-complexity coding solutions for the traditional MAC channel (e.g. \cite{rimoldi1996rate}) assume coordination between the users. Due to the gigantic number of users we assume them to be symmetric, i.e. the users use the same codes and equal powers. 

\subsection{Transmission}
Let us list the main features of the transmission process: 
\begin{itemize}
\item the frame is split into $V$ slots of size $n = N/V$ channel uses;
\item the user chooses a message $\omega$, then encodes it and obtains a codeword $\mathbf{x}(\omega)$ of length $n$; 
\item users repeat their codewords in multiple slots. The repetition count distribution is the same for all the users. By $D(r)$ we denote the probability, that $r$ replicas will be sent;
\item the number of repetitions $r$ and the $r$ slots in which to send are chosen based on the message $\omega$ (see \cite{Vem2017}): as $\omega$ is distributed uniformly on $[M]$ the slots are chosen uniformly at random (without repetitions) from $V$ existing slots.   
\end{itemize}

\subsection{Joint decoding within a slot}

Let us first note, that in order to obtain an achievability bound for the whole scheme we do not restrict the complexity of the slot decoding and want to use randomly generated codebook from \cite{polyanskiy2017perspective}. 

Consider a particular slot, w.l.o.g let it be the first slot. Let $\mathbf{y}_1$ be the received signal of $n$ channel uses. We also assume, that $K_1$ users transmit in the first slot. Recall, that $k$ denotes the number of information bits to be sent by each user and $P$ is the average transmit power. The random coding bound \cite{polyanskiy2017perspective} states, that in random Gaussian ensemble there exists a codebook $\mathcal{C}'$, such that
\begin{IEEEeqnarray*}{LL}
&\frac{1}{K} \sum\limits_{i=1}^{K} \Pr(W_i \not\in \mathcal{L}(\mathbf{{y}}_1) | \mathcal{C}') = \Pr(W_1 \not\in \mathcal{L}(\mathbf{{y}}_1) | \mathcal{C}')\\
&\leq p(n, k, P, K_1) \triangleq \mathbb{E}_\mathcal{C}[\Pr(W_1 \not\in \mathcal{L}(\mathbf{{y}}_1) | \mathcal{C}, K_1)].
\end{IEEEeqnarray*}

The first equality holds due to the symmetry of users (it is enough to consider the probability that a particular user's message is not in the decoded list). The last expectation is taken over the Gaussian ensemble. 

Here we emphasize \textit{the main problem} -- the bound assumes the number of active users ($K_1$) to be known. Due to the randomness of $T$-fold IRSA the number of users transmitting in a particular slot is a random variable. We note, that due to the short slot length it is not possible to estimate (e.g. by energy) the number of users with satisfactory probability of error. Thus we need to perform a blind decoding. One another problem is that a codebook $\mathcal{C}'$ is constructed for a particular number of users, but we need a codebook, that can resolve collisions of order $K_1 \in \{1, \ldots, T\}$.  

To deal with these problems let us change the decoder. Let $S \subset [M]$, $|S| = K_1$, denote the set of messages, that were transmitted. The decoding rule is as follows
\[
\hat{S} = \argmin\limits_{\hat{S} \subset [M], |\hat{S}| \leq T} || \mathbf{y_1} - c(\hat{S}) ||_2^2,
\]
where $\hat{S}$ is an estimate of $S$, $c(\hat{S}) = \sum\nolimits_{j \in \hat{S}} \mathbf{x}_j$. Recall, that $T$ is the maximal collision order, that can be resolved.



\begin{theorem}\label{theor_av}
Fix $P' < P$ and $T$, then the average (over random Gaussian ensemble) per user error probabilities can be calculated as follows ($K_1 = 1, \ldots, T$)
\[
\mathbb{E}_\mathcal{C} \left[ \Pr(W_1 \not\in \mathcal{L}(\mathbf{{y}}_1) | \mathcal{C}, K_1) \right] \leq \sum\limits_{t=1}^{K_1} \frac{t}{K_1} p_t(K_1, T) + p_0(K_1),
\]
where
\begin{eqnarray*}
p_0(K_1) &\leq& \frac{\binom{K_1}{2}}{M} + K_1 \Pr\left[ \frac{1}{n} \sum\limits_{j=1}^{n} Z_j^2 > \frac{P}{P'} \right] \\
p_t(K_1, T) &=& \sum\limits_{\hat{t} = 0}^{T-K_1+t}e^{-n E(t, \hat{t})}\\
E(t, \hat{t}) &=& \max\limits_{0\leq \rho, \rho_1 \leq 1, \lambda > 0} -\rho\rho_1 t R_1 - \rho_1 R_2 + E_0\\
E_0 &=& \rho_1 a + \frac{1}{2} \log(1-2b\rho_1)\\
a &=& \frac{\rho}{2} \log (1+2P'\hat{t}\lambda) + \frac{1}{2} \log (1+2P't\mu)\\
b &=& \rho \lambda - \frac{\mu}{1+2P't\mu}, \mu = \frac{\rho \lambda}{1+2P'\hat{t}\lambda} \\
R_1 &=& \frac{1}{n} \log M - \frac{1}{n \hat{t}} \log(\hat{t}!)\\ 
R_2 &=& \log \binom{K_1}{t}.
\end{eqnarray*}
\end{theorem}
\begin{IEEEproof}

In the proof we just emphasize the difference in comparison to \cite{polyanskiy2017perspective}. Let us first define the decoding error condition
\[
\Vert \mathbf{y}_1 - c(S) \Vert_{2}^{2} > \Vert \mathbf{y}_1 - c(\hat{S}) \Vert_{2}^{2}.
\]
Let us introduce the additional notation. Let $S_0 = S \backslash \hat{S}$ and $\hat{S}_0 = \hat{S} \backslash S$. We are interested in estimating the probability $\Pr[|S_0| = t]$. In this case (due to the decoding rule) $0 \leq |\hat{S}_0| \leq T-K_1+t$. The notation is illustrated in \Fig{fig:notation}. We can rewrite the error condition as follows
\[
\Vert \mathbf{z}_1 \Vert_{2}^{2} > \Vert c(S_{0}) - c(\hat{S_{0}}) + \mathbf{z}_1 \Vert_{2}^{2},
\]
where $\mathbf{z}_1$ is a noise vector.

 \begin{figure}[t]
   \centering
   \includegraphics[width=0.7\linewidth]{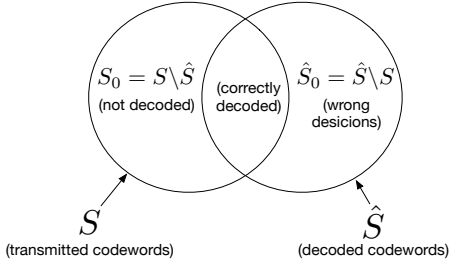}
   \caption{Notation}
   \label{fig:notation}
 \end{figure}

Let $|\hat{S}_0| = \hat{t}$, let us introduce the events
\[
F(S_0, \hat{S}_0) \triangleq \left\{ \Vert \mathbf{z}_1 \Vert_{2}^{2} > \Vert c(S_{0}) - c(\hat{S_{0}}) + \mathbf{z}_1 \Vert_{2}^{2} \right\},
\]
and
\[
F(S_0) \triangleq \mathop{\cup}\limits_{\hat{S}_0} F(S_0, \hat{S}_0).
\]

Using Chernoff bound (let $\lambda >0$)
\[
\Pr\left[ F(S_0, \hat{S}_0) | S_0, \hat{S}_0, c(S_0), \mathbf{z}_1 \right] \leq A(t, \hat{t}),
\]
where
\[
A(t, \hat{t}) = e^{\lambda \Vert \mathbf{z}_1 \Vert^2_2} \frac{e^\frac{-\lambda \Vert c(S_0) + \mathbf{z}_1\Vert_{2}^{2}}{1+2\hat{t} \lambda P'}} {(1+2\hat{t} \lambda P')^{n/2}}
\]

Thus,
\[
\Pr\left[ F(S_0) | S_0, c(S_0), \mathbf{z}_1 \right] \leq \sum\limits_{\hat{t} = 0}^{T-K_1+t} \binom{M-K_1}{\hat{t}} A(t, \hat{t}). 
\]

Averaging over $c(S_0)$ and $\mathbf{z}_1$ is done exactly the same as in \cite{polyanskiy2017perspective}.
\end{IEEEproof}

We now only need to choose a codebook, that is good for all the collision orders up to $T$. Let us formulate a statement.

\begin{statement}\label{stat:code}
Let us choose positive values $\alpha_i$, $i = 1, \ldots, T$, such that $\sum\nolimits_{i=1}^T \alpha_i < 1$. In a random Gaussian ensemble there exists a codebook $\mathcal{C}^*$, such that the following inequalities hold for all $K_1 = 1, \ldots, T$
\begin{IEEEeqnarray*}{LL}
&\Pr(W_1 \not\in \mathcal{L}(\mathbf{{y}}_1) | \mathcal{C^*}, K_1) \leq \tilde{p}(n, k, P, T, K_1) \\
& \triangleq \frac{1}{\alpha_i}\mathbb{E}_\mathcal{C} \left[ \Pr(W_1 \not\in \mathcal{L}(\mathbf{{y}}) | \mathcal{C}, K_1) \right]
\end{IEEEeqnarray*}
\end{statement}\label{stat::code}
\begin{IEEEproof}
We need to estimate a probability of a bad code -- a code for which the inequalities does not hold at least for one $K_1$. By applying the Markov's inequality and a union bound we see, that this probability is upper bounded by $\sum\limits_{i=1}^T \alpha_i < 1$.
\end{IEEEproof}

\subsection{SIC decoder}

Decoding algorithm is based on successive interference cancellation approach. At each step the algorithm selects a slot from the set of unresolved slots. Then a joint decoding algorithm (see above) is applied for this slot to extract the user's messages from the received signal. As a result some messages are decoded successfully, some of the messages are decoded incorrectly. Then all successfully decoded messages are removed from other slots (if the message was transmitted by user more than once) and the slot itself is marked as resolved. We note, that we can always find where the replicas were transmitted as these positions are chosen based on the data (so in contrast to \cite{Liva} we do not need to store the pointers). The algorithm stops when the set of unresolved slots is empty.

\begin{remark}
We note, that during the slot decoding the errors may occur, i.e. some of the packets (codewords) may be decoded incorrectly. In what follows we assume, that we can always detect the error packets (the packets include control information). 
\end{remark}

\section{Density evolution}
The transmission and decoding processes can be described with the use of a bipartite graph, which is called the Tanner graph \cite{tanner1981recursive}. The vertex set of the graph consists of the set of user nodes $U = \{u_1, u_2, \ldots, u_K\}$ which correspond to the set of users and the set of slot nodes $C = \{c_1, c_2, \ldots, c_V\}$ which correspond to signals received in slots. The user node $u_i$ and the slot node $c_j$ are connected with an edge if and only if the $i$-th users transmitted a packet in the $j$-th slot.

\begin{figure}[t]
\centering
\includegraphics[width=0.6\linewidth]{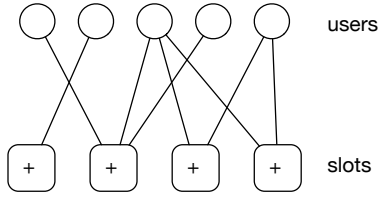}
\caption{Tanner graph representation}
\label{fig:tanner_repr}
\end{figure}

Let $L(x) = \sum\nolimits_{i}L_i x^i$ and $\lambda(x) = \sum\nolimits_i \lambda_i x^{i-1}$ denote the user node degree distributions from node and edge degree perspective, respectively. We recall (see e.g. \cite{richardson2008modern}), that $L_i$ and $\lambda_i$ denote respectively the fractions of user nodes of degree $i$ and the fraction of edges incident to user nodes of degree $i$. Also recall, that $\lambda(x) = L'(x)/L'(1)$. In our case $L_i = D(i)$. Analogously, let $R(x) = \sum\nolimits_{i}R_i x^i$ and $\rho(x) = \sum\nolimits_i \rho_i x^{i-1}$ denote the slot node degree distributions from node and edge degree perspective, respectively.

Let $G = K/V$. Let us consider the $j$-th slot. Each user chooses this slot for transmission independently with probability $\frac{L'(1)}{V} = \frac{G L'(1)}{K}$. Thus, the slot node distribution (from node perspective) is $\text{Bino}\left( K, \frac{G L'(1)}{K} \right)$. In the limit $K \to \infty$ this distribution becomes a Poisson distribution. In what follows we use $R(x) = \rho(x) = e^{-G L'(1) (1-x)}$.

Similar to \cite{Liva, Pfister} we consider the ensemble of Tanner graphs $\mathcal{G}(K;V; \lambda(x); \rho(x))$ corresponding to the multiple-access scheme with $K$ users, $V$ slots, and the degree distributions $\lambda(x)$ and $\rho(x)$. We are interested in the decoding performance averaged over the ensemble $\mathcal{G}(K;V; \lambda(x); \rho(x))$ in the limit as $K$,$V \to \infty$.

Our main goal is to minimize the required energy-per-bit $E_b/N_0$. For this purpose we use a DE method, which helps us to choose the system parameters. We note, that the modification of DE for the case of multi-packet reception can be found in \cite{Liva} (see the appendix), but here we apply it for the noisy channel and combined with a finite length random coding bound. The  major difference of our approach is that we
\begin{itemize}
    \item take into account a noisy channel (AWGN channel to be precise);
    \item take into account finite length effect as the slots have small length;
    \item take into account a transmit energy -- \textit{energy efficiency} is our main optimization criterion. Assume we use a strategy with $L(x) = x^2$. In this case we spent $2$ times more energy while transmitting in comparison to $L(x) = x$ strategy;
\end{itemize}

Recall, that $n$ is a slot length, $k$ is the number of information bits to be sent by each user, $P$ is the average transmit power (linear scale). Let us fix the maximal number of iterations $\ell$ and $L(x)$. Then the average energy per information bit can be calculated as follows
\[
\frac{E_b}{N_0} = \frac{n P L'(1)}{2k},
\]
we note, that $L'(1)$ is actually the average number of transmissions and $L'(x)$ means a derivative with respect to $x$.

Now let us write the density evolution rules. We assume, that a code $\mathcal{C}^*$ constructed in accordance to Statement~\ref{stat::code} is used in the system. By $x_l$ and $y_l$ we denote the probability that an outgoing message from the user node and slot node, respectively, are erased during the $l$-th iteration. We start with initial condition $x_0 = 1$, which means, that the user messages are erased at the beginning and we observe only the noisy signal sums in slots. 

\begin{eqnarray*}
y_{l+1} &=& 1 - \rho(1-x_l) \sum\limits_{t=0}^{T-1} \left( 1 - \tilde p(n, k, P, T, t+1)\right) \\ &\times& \frac{\left( G L'(1) x_l \right)^t}{t!} \\
x_{l} &=& \lambda(y_l), \:\: 1 \leq l < \ell.\\
x_{\ell} &=& L(y_{\ell}).
\end{eqnarray*}

\begin{IEEEproof}

Consider the $l$-th iteration. We want to calculate the erasure probability of the outgoing message $y_{l+1}$ based on incoming messages (with erasure probabilities $x_l$). The probability can be calculated as follows (recall, that $\rho_r$ is the probability, that the outgoing edge is connected to a slot node of degree $r$)

\begin{eqnarray*}
y_{l+1} &=& 1 -\sum\limits_{r = 1}^{r_{\max}} \rho_r \Big[ \sum\limits_{t=0}^{\min(r,T)-1} \left( 1 - \tilde p(n, k, P, T, t+1)\right)  \\
&\times& \binom{r-1}{t} x_l^t (1-x_l)^{r-1-t} \Big]
\end{eqnarray*}

By changing the summation order we obtain the needed result.

\end{IEEEproof}

\begin{remark}
We note, that 
\[
\lim\limits_{\ell \to \infty} x_\ell > 0.
\]
because of finite length effects in the slot. So in what follows we do not consider infinite number of iterations and fix $\ell$.
\end{remark}

\begin{figure}[t]
\centering
\includegraphics[width=\columnwidth]{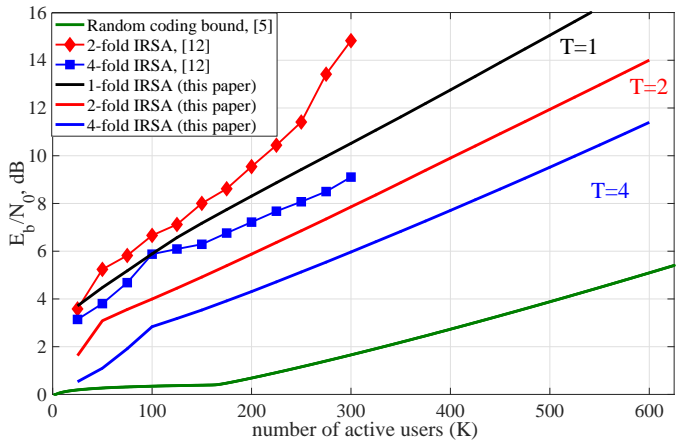}
\caption{Minimal $E_b/N_0$ required to achieve less than 5\% error rate as a user count function.} 
\label{fig:t_2}
\end{figure}
\section{Numerical results}

We choose the same system parameters as in \cite{polyanskiy2017perspective, OrdentlichPolyanskiy2017, Vem2017} for honest comparison, i.e. $N=30000$, $k=100$, $p^* = 0.05$ (maximal allowed $p_e$).

\subsection{Optimization procedure}
The goal is to find $n$ and corresponding $G = K / V = n K / N$ as well as the polynomial $L(x)$ in order to minimize the $E_b/N_0$ under the maximum allowed per user error probability $p_e$.
$$
\left\{L\left(x\right), n\right\} = \argmin_{L\left(x\right), n}\left(\frac{E_b}{N_0}: p_e \leq p^* \right)
$$

The optimization procedure is conducted separately for every $K$ and consists of two sub-procedures. The first one is to find a local minimum of $p_e$ with respect to $L(x)$ and $n$ under the following constraints.
$$\quad L(1) = 1,\quad n > 0, \quad L_i \geq 0 \quad \forall i = 1, \ldots, d_{\max},$$
where $d_{\max}$ is the maximum polynomial degree allowed. This sub-procedure is performed with fixed $E_b/N_0$. In order to find a global minimum of $p_e$ one need to run multiple optimization procedures starting from different random initial points within constraints.

We expect $p_e$ to be a monotonic function of $E_b/N_0$ and use a binary search procedure to find the minimal $E_b/N_0$ where the $p_e \leq p^{\star}$ holds.

Optimal node degree distribution, slot count $K/G$ and $E_b/N_0$ are shown in tables \ref{DE_T1}, \ref{DE_T2} and \ref{DE_T4} for $T=1$, $2$, $4$ respectively for $10$ DE iterations.

The numerical results show that $L(x)$ behaves smoothly when varying the number of users. This means that a global minimum is found at every optimization point. Note, that the error probability has multiple local minima, because the slot count changes sharply at several points.

\begin{table}
\vskip 0.1cm
\caption {Optimal node degree distribution, slot count $K/G$ and $E_b/N_0$ (dB) for $T=1$, $10$ iterations} 
\label{DE_T1}
\centering
\begin{tabular}{|r|l|r|r|r|}
\hline
$K$ & $L(x)$ & $G$ & $K/G$ & $E_b/N_0$ \\
\hline
25   & $0.0928x + 0.9072x^2$             & 0.364 &  68.67 &  3.71 \\ 
50   & $1.0000x^2$                       & 0.455 & 109.90 &  4.48 \\ 
100  & $1.0000x^2$                       & 0.490 & 204.06 &  5.89 \\ 
150  & $0.6211x^2 + 0.3789x^3$           & 0.604 & 248.25 &  7.17 \\ 
200  & $0.4781x^2 + 0.5219x^3$           & 0.643 & 310.91 &  8.31 \\ 
250  & $0.0706x + 0.2011x^2 + 0.7283x^3$ & 0.674 & 370.71 &  9.42 \\ 
300  & $0.1297x + 0.8703x^3$             & 0.693 & 432.89 & 10.52 \\ 
350  & $0.1234x + 0.8766x^3$             & 0.698 & 501.62 & 11.62 \\ 
400  & $0.1184x + 0.8816x^3$             & 0.702 & 570.22 & 12.76 \\ 
450  & $0.1247x + 0.7991x^3 + 0.0763x^4$ & 0.710 & 633.93 & 13.90 \\ 
500  & $0.1396x + 0.6716x^3 + 0.1889x^4$ & 0.720 & 694.53 & 15.05 \\ 
550  & $0.1474x + 0.5906x^3 + 0.2620x^4$ & 0.726 & 757.85 & 16.20 \\ 
600  & $0.1549x + 0.5239x^3 + 0.3212x^4$ & 0.730 & 821.46 & 17.37 \\ 
\hline
\end{tabular}
\end{table}

\begin{table}
\caption {Optimal node degree distribution, slot count $K/G$ and $E_b/N_0$ (dB) for $T=2$, $10$ iterations} 
\label{DE_T2}
\centering
\begin{tabular}{|r|l|r|r|r|}
\hline
$K$& $L(x)$ & $G$ & $K/G$ & $E_b/N_0$ \\
\hline
 25  & $1.0000x$                         & 0.317 &  78.85 &  1.63 \\ 
 50  & $0.4443x + 0.5557x^2$             & 0.830 &  60.28 &  3.09 \\ 
100  & $0.2099x + 0.7901x^2$             & 1.283 &  77.95 &  4.00 \\ 
150  & $0.1680x + 0.8320x^2$             & 1.358 & 110.49 &  4.92 \\ 
200  & $0.1382x + 0.8618x^2$             & 1.395 & 143.38 &  5.88 \\ 
250  & $0.1205x + 0.8795x^2$             & 1.419 & 176.18 &  6.86 \\ 
300  & $0.1008x + 0.8992x^2$             & 1.435 & 209.06 &  7.86 \\ 
350  & $0.0895x + 0.9105x^2$             & 1.448 & 241.74 &  8.87 \\ 
400  & $0.1206x + 0.7609x^2 + 0.1185x^3$ & 1.485 & 269.35 &  9.90 \\ 
450  & $0.1609x + 0.5953x^2 + 0.2438x^3$ & 1.518 & 296.52 & 10.92 \\ 
500  & $0.1867x + 0.4916x^2 + 0.3217x^3$ & 1.537 & 325.41 & 11.94 \\ 
550  & $0.2061x + 0.4160x^2 + 0.3779x^3$ & 1.550 & 354.95 & 12.97 \\ 
600  & $0.2184x + 0.3639x^2 + 0.4177x^3$ & 1.558 & 385.14 & 14.00 \\
\hline
\end{tabular}
\end{table}

\begin{table}
\centering
\caption {Optimal node degree distribution, slot count $K/G$ and $E_b/N_0$ (dB) for $T=4$, $10$ iterations} 
\label{DE_T4}
\begin{tabular}{|r|l|r|r|r|}
\hline
$K$ & $L(x)$ & $G$ & $K/G$ & $E_b/N_0$\\
\hline
 25  & $1.0000x$             & 1.017 &  24.59 &  0.53 \\ 
 50  & $1.0000x$             & 1.218 &  41.04 &  1.10 \\ 
100  & $0.5781x + 0.4219x^2$ & 2.317 &  43.16 &  2.84 \\ 
150  & $0.3987x + 0.6013x^2$ & 3.053 &  49.14 &  3.54 \\ 
200  & $0.3764x + 0.6236x^2$ & 3.126 &  63.99 &  4.31 \\ 
250  & $0.3620x + 0.6380x^2$ & 3.165 &  78.98 &  5.13 \\ 
300  & $0.3538x + 0.6462x^2$ & 3.194 &  93.93 &  5.97 \\ 
350  & $0.3479x + 0.6521x^2$ & 3.216 & 108.84 &  6.83 \\ 
400  & $0.3416x + 0.6584x^2$ & 3.232 & 123.75 &  7.71 \\ 
450  & $0.3349x + 0.6651x^2$ & 3.245 & 138.69 &  8.60 \\ 
500  & $0.3302x + 0.6698x^2$ & 3.255 & 153.60 &  9.52 \\ 
550  & $0.3269x + 0.6731x^2$ & 3.265 & 168.48 & 10.45 \\ 
600  & $0.3225x + 0.6775x^2$ & 3.271 & 183.42 & 11.40 \\
\hline
\end{tabular}
\end{table}
\subsection{Simulation results with fixed number of active users}
Interference cancellation algorithm was tested via Gaussian MAC Monte Carlo simulations. The result of a single run is a set of slots and the number of simultaneous transmissions (or collision index) for each slot. Each user selects the number of transmissions in accordance to $L(x)$ and then selects particular non-coinciding slots from uniform distribution during each run. The same $E_b/N_0$ is assumed for all slots.

The decoding is done in accordance to SIC algorithm. The only thing we need to explain is how we resolve the collisions. Error probability is set to $1$ if the number of simultaneous transmissions within some slot exceeds the threshold ($T\in\left\{1, 2, 4\right\}$). If the order of collision is less or equal to $T$, then the error probability is calculated independently for each transmitted message in a slot in accordance to finite length random coding bound (see Statement~\ref{stat:code}).

Monte-Carlo validation shows that developed density evolution method with $10$ iterations predicts the performance pretty well, i.e. the simulated error rate does not exceed $5$\% if the minimal $E_b/N_0$ is increased by $0.15$--$0.2$ dB.
\subsection{Random number of users}
For now, suppose the number of users to be a random value and suppose that this random value has a Poisson distribution with a mean value $K$. Let us solve the same task, i.e. find minimal $E_b/N_0$ which can guarantee the \emph{mean} error rate (under Poisson distribution) to be not higher than $p^*$. In order to choose optimized polynomial $L(x)$ for this case we again use density evolution method. The difference in comparison to the previous case is as follows. We need to average the probability of error over the number of users. 



\begin{figure}[t]
\centering
\includegraphics[width=\columnwidth]{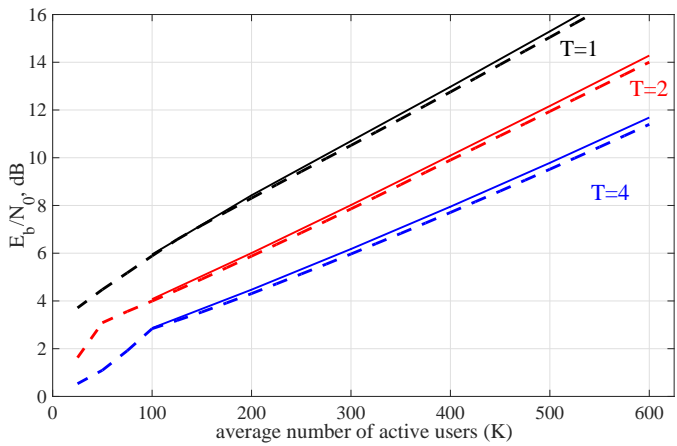}
\caption{Minimal $E_b/N_0$ required to achieve $5$\% error rate as a user count function for random user count (solid line) and deterministic user count (dashed line).}
\label{fig:t_2_random}
\end{figure}

With the density evolution method described above we can find optimized polynomials $L(x)$ and minimal $E_b/N_0$ values (in order to guarantee $p_e \leq p^*$) for the random user count case. In \Fig{fig:t_2_random} we present the comparison of required $E_b/N_0$ values for deterministic and random cases. As previously the polynomials were optimized separately for different values of $K$, so for each $K$ we have different polynomial. 

We see that the difference is not big. We also note, that the polynomials $L\left(x\right)$ do not differ significantly when shifting between random and deterministic user count. 


\section{Conclusion}
In this paper we derived and presented achievability bounds for T-fold IRSA scheme in the Gaussian MAC. In order to do this we used density evolution method in combination with random coding bound proposed by Y.~Polyanskiy. To the best of the authors' knowledge, the bounds presented in this paper are the best achievability bounds for low-complexity random coding schemes, which can be used in this channel. 

In order to finalize the scheme it is important to construct user codes with the performance close to the random coding bound. Due to the fact that we want to construct the random access scheme the users have to utilize the same codebook. Thus, the task of constructing same codebook codes with low complexity decoding (say, same codebook LDPC codes) is very important. One another research direction is considering of a fading MAC. It will be very interesting to generalize the results presented here for the fading scenario.


\bibliographystyle{IEEEtran}
\balance
\bibliography{bibl}
\end{document}